\documentclass[preprint2]{emulateapj}

\usepackage[small,normal,bf,up]{caption}
\renewcommand{\captionfont}{\small}

\shorttitle{The Metallicity Gradient of a $z=2$ Galaxy}

\shortauthors{Jones et al.}

\begin{document}


\title{Measurement of a Metallicity Gradient in a $z=2$ Galaxy: \\ Implications for Inside-Out Assembly Histories}

\author{Tucker Jones and Richard Ellis}
\affil{Astronomy Department, California Institute of Technology, MC 249-17, Pasadena, CA 91125, USA}
\author{Eric Jullo}
\affil{Jet Propulsion Laboratory, Caltech, MS 169-327, Pasadena CA 91125, USA}
\and
\author{Johan Richard}
\affil{Institute for Computational Cosmology, Department of Physics, Durham University, South Road, Durham, DH1 3LE, UK}



\begin{abstract}

We present near-infrared imaging spectroscopy of the strongly-lensed $z=2.00$ galaxy SDSS J120601.69+514227.8 (`the Clone arc'). Using OSIRIS on the Keck 2 telescope with laser guide star adaptive optics, we achieve resolved spectroscopy with 0.20 arcsecond FWHM resolution in the diagnostic emission lines [O{\sc iii}], H$\alpha$, and [N{\sc ii}]. The lensing magnification allows us to map the velocity and star formation from H$\alpha$ emission at a physical resolution of $\simeq 300$ pc in the galaxy source plane. With an integrated star formation rate of $\simeq 50 M_{\odot}$ yr$^{-1}$, the galaxy is typical of sources similarly studied at this epoch. It is dispersion-dominated with a velocity gradient of $\simeq\pm 80$ km s$^{-1}$ and average dispersion $\bar{\sigma} = 85$ km s$^{-1}$; the dynamical mass is $2.4 \times 10^{10} M_{\odot}$ within a half-light radius of 2.9 kpc. Robust detection of [N{\sc ii}] emission across the entire OSIRIS field of view enables us to trace the gas-phase metallicity distribution with 500 pc resolution. We find a strong radial gradient in both the [N{\sc ii}]/H$\alpha$ and [O{\sc iii}]/H$\alpha$ ratios indicating a metallicity gradient of $-0.27 \pm 0.05$ dex kpc$^{-1}$ with central metallicity close to solar.  We demonstrate that the gradient is seen independently in two multiple images. While the physical gradient is considerably steeper than that observed in local galaxies, in terms of the effective radius at that epoch, the gradient is similar. This suggests that subsequent growth occurs in an inside-out manner with the inner metallicity gradient diminished over time due to radial mixing and enrichment from star formation.

\end{abstract}



\section{Introduction}

Considerable progress is being made with integral field unit (IFU) spectrographs on large telescopes in
unraveling the internal properties of star-forming galaxies over $1<z<5$. During this period, the
mass assembly of galaxies proceeds at a rapid pace and such measures promise valuable
insight into the physical mechanisms by which young galaxies grow. Early effort has focused on the dynamical properties
(e.g. \citealt{Forster09, Law09, Jones10, Gnerucci10}). 
These studies have revealed systems which are dispersion-dominated with varying degrees of ordered rotation consistent
with sources which may develop stable disks and central stellar bulges. The turbulent motions may be associated 
with vigorous star-formation driven by cold molecular gas accreted along filaments
from the nearby intergalactic medium \citep{Tacconi10}. The highest resolution data, secured by coupling
the angular magnification of strong gravitational lensing with laser-assisted guide star adaptive optics (LGSAO),
has revealed the sizes of star-forming regions in $z\simeq$2-3 galaxies suggesting that star formation is primarily 
triggered by gravitational instability rather than external mergers \citep{Stark08, Jones10, Swinbank10}.

Attention is now shifting to understanding the evolution of the mass-metallicity relation \citep{Tremonti04}. 
Metallicity is a key parameter which gauges the baryonic material already converted into stars thus offering 
insight into feedback processes proposed to regulate star formation. The trend is usually explained via
star-formation driven outflows, e.g.\ from energetic supernovae,  which have a larger effect in low mass galaxies 
with weaker gravitational potentials. However, other effects may be responsible at high redshifts where star 
formation timescales and feedback processes and their mass dependences differ. Much effort has been invested 
in measuring evolution in the relation using the integrated light from samples beyond 
$z\simeq2$ \citep{Erb06, Halliday08, Mannucci09, Hayashi09}. These pioneering surveys have shown that 
metallicity decreases with redshift for a fixed stellar mass. 

A logical next step in developing a picture of the assembly history of disk galaxies would be to extend the kinematic 
progress made with IFU spectrographs to {\it resolved measures of the gas phase metallicity}. 
This would enable the tracking of metallicity gradients as a function of redshift and provide a direct test of 
models of star formation and assembly induced by cold gas inflows \citep{Brooks07}. However, even with 
LGSAO, the spatial sampling for representative  $z\simeq$2-3 galaxies is $>$1 kpc and insufficient for
this purpose. Prior to large telescopes with improved diffraction-limited performance, gravitationally-lensed sources 
offer  a natural starting point (c.f. \citealt{Stark08}). In this Letter we examine the practicality of this approach via 
a detailed study of the gravitationally-lensed arc SDSS J120602.09+51422, referred to as the `Clone'  arc \citep{Lin09}. 
Hubble Space Telescope (HST) images and spectroscopic follow-up by \cite{Lin09} confirm this to be a $z$=2.001 star-forming galaxy with intrinsic half-light radius $r_h = 2.9$ kpc and SFR $\simeq\,50 M_\odot$ yr$^{-1}$,
magnified an areal factor $\simeq$30 by a foreground $z$=0.422 group of galaxies. Further considerations 
of the lensing configuration have been presented by \cite{Vegetti10}. The object was selected
for detailed study after considering an integrated near-infrared emission line spectrum using NIRSPEC 
on Keck 2 (Hainline et al 2009).

Throughout this paper we adopt a $\Lambda$\,CDM cosmology with $H_0$= 70 km s$^{-1}$
Mpc$^{-1}$, $\Omega_M$=0.30 and $\Omega_\Lambda$=0.70. At $z$=2.00, 0.1 arcsec corresponds to 840 pc and the age of the universe was 3.2 Gyr. 
All magnitudes are in the AB system.

\section{Observations and Data Reduction}

\begin{figure*}
\plotone{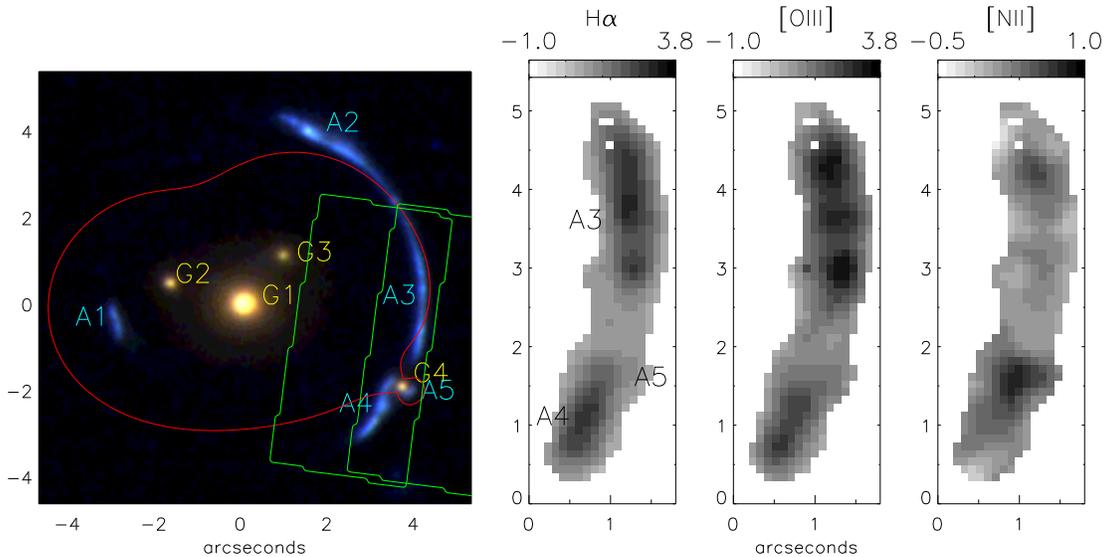}
\caption{\label{fig:cl} (Left) Hubble Space Telescope color composite image of the Clone arc. A1-5 represent
multiple images of the $z$=2.00 source; the critical curve is shown in red. The OSIRIS pointings are indicated by 
the two green rectangles offset by 1.8 arcseconds in the East-West direction. Foreground lensing galaxies are 
labeled as G1-4. (Right) Distribution of the key emission line fluxes in multiple images A3-5 in units of 
$10^{-18}$ erg/s/cm$^2$ (see text for discussion of optimum sampling). North is up and East is to the left. The critical line 
passes through the A3 and A4 components such that a small part of the arc (including A5) is imaged 4 times in the OSIRIS field. 
The remainder of A3/A4 is imaged twice in the OSIRIS field.}
\end{figure*}

Observations of SDSS J120601.69+514227.8 (the Clone) were taken with the OH-Suppressing Infra-Red 
Imaging Spectrograph (OSIRIS; \citealt{Larkin06}) and the LGSAO 
system on the Keck 2 telescope on 19 May 2010. The weather was clear with strong winds 
and the seeing was 1.5 arcsecond FWHM.  Nonetheless, tip/tilt correction yielded
a respectable Strehl of 0.15 at $2\, \mu$m.
We took exposures of 900 seconds 
each, dithering by 1.8 arcseconds in order to achieve good sky subtraction while keeping the arc 
within the OSIRIS field of view. We obtained 4 exposures in the Kn1 band (covering H$\alpha$ and [N{\sc ii}]), 2 
exposures in Hn1 ([O{\sc iii}]) and 3 exposures in Zn4 ([O{\sc ii}]).
The physical location of the OSIRIS field is shown on a HST color composite in Figure~\ref{fig:cl}.

Dark and bias subtraction, spectral extraction, wavelength calibration and construction of the relevant
data cubes was accomplished using the OSIRIS data reduction pipeline (ODRP; \citealt{Larkin06}). Sky subtraction 
was done with the IDL code described in \cite{Davies07} using adjacent pairs of images as sky 
reference frames. The final data cubes were stacked using a clipped mean and show strong [O{\sc iii}], H$\alpha$, 
and [N{\sc ii}] emission. The brightest individual pixels have a signal-to-noise of $\simeq$10 in H$\alpha$, 
5 in [O{\sc iii}], and 2 in [N{\sc ii}]. The [O{\sc ii}] doublet was not reliably detected and H$\beta$ lay in
a region of low atmospheric transmission.

The infrared standard star FS 133 was observed with the same instrument configuration.
Precise {\sl JHK} UKIRT photometry from \cite{Hawarden01} was matched to the G8{\sc v} stellar spectrum from 
\cite{Pickles98} to determine the conversion from ADU to flux as a function of wavelength. We tested 
this calibration by comparing the ratio of observed flux (in ADU\,s$^{-1}$) to the reference spectrum 
(in erg\,s$^{-1}$\,cm$^{-2}$) with an atmospheric transmission model generated by the ATRAN program 
\citep{Lord92}. The data reproduced all major atmospheric absorption features to better than 10 \% accuracy. 
We checked further for systematic errors by comparing the flux ratio of bright emission lines in the integrated 
OSIRIS spectrum of image A3 (Figure~\ref{fig:cl}) with independent results from NIRSPEC \citep{Hainline09}. 
The results were consistent to within the uncertainties (6\%). Overall we deduce
a systematic uncertainty in flux ratios of $\simeq 10$\,\% for lines observed with different instrumental 
configurations (such as [O{\sc iii}]/H$\alpha$), and negligible systematic error for lines observed simultaneously 
([N{\sc ii}]/H$\alpha$).

\section{Gravitational lens model}\label{sec:lensmodel}

\begin{table*}
\begin{tabular}{cccccccc}
\tableline\tableline
        Component & $\theta_E$ & $\theta$ & $q$ & $x_c$ & $y_c$ \\
\tableline
        G1 & $3.0\pm0.3$ & $-81\pm6$ & $0.64\pm0.05$ & $0.11\pm0.08$ & $0.0\pm0.08$ \\
        G2 & $0.5\pm0.3$ &  \\
        G3 & $0.3\pm0.2$ &  \\
        G4 & $r_t = 0.68''$ & $M_{tot} = 1.4\pm0.2\times 10^{10} \mathrm{M}_\odot$ \\
	Shear & $\Gamma_{sh} = 0.05\pm0.02$ & $\theta_{sh} = 63\pm21$ \\
\tableline
\end{tabular}
\caption{Best fit parameter values and 1$\sigma$ uncertainties for each component of our lens model. Components G2, G3 and G4 are assumed to be spherical mass distributions located at the galaxy centroids. The central position of G1 is in arcseconds relative to the galaxy centroid.}
\end{table*}

Our OSIRIS field samples several multiple images of the $z$=2.00 galaxy (A3-5; Figure~\ref{fig:cl}) enabling us, 
in principle, to derive the source plane properties independently via each image.  However, accurate modeling 
of the foreground gravitational lens is necessary to realize this advantage. We 
use the {\sc lenstool} program \citep{Kneib93, Jullo07} to parameterize the mass distribution of the lens, 
using the image positions as constraints.

Our lens model comprises 1 group-scale dark matter halo, 3
galaxy-scale dark matter halos associated with G2, G3, and G4, and an
external shear component. Galaxies G1, G2, and G3 are parameterized
with a singular isothermal ellipsoid (SIE). G4 is
parameterized with a dual Pseudo-Isothermal elliptical mass
distribution (dPIE; \citealt{Eliasdottir07}) with an assumed
tidal radius of 0.68'' \citep{Vegetti10}.
The fit is very good, with a mean
residual error of 0.2'' in the image plane and 0.06'' in the source
plane.  The recovered parameter values for this model are reproduced
in Table 1. We find typical linear magnification factors of $1.05\times12.5$ (orthogonal and parallel 
to the arc) for image A3 and $1.01\times6.7$ for A4 with an uncertainty of 5\,\%. The areal magnification, 
$28.1 \pm 1.4$, is in good agreement with \cite{Lin09}.

{\sc Lenstool} defines the necessary transformations to the source (intrinsic) plane for each image, 
giving the source plane position for each spatial pixel of OSIRIS. We use the {\sc IDL} functions 
{\sc triangulate} and {\sc trigrid} to interpolate the irregularly-sampled source plane data onto a uniform grid. The 
source plane resolution of $0.15 \times 1.6$ kpc FWHM was measured by reconstructing the tip/tilt reference star in the 
source plane at various locations. Since we will smooth the data to improve the signal/noise we note here that the 
resolution increases to $0.3 \times 2.3$ kpc and $0.5 \times 3.0$ kpc  when the image plane data is convolved 
with a Gaussian of FWHM $= 0.25$ and 0.40 arcseconds, respectively.

\section{Kinematic Properties}

\begin{figure*}
\plotone{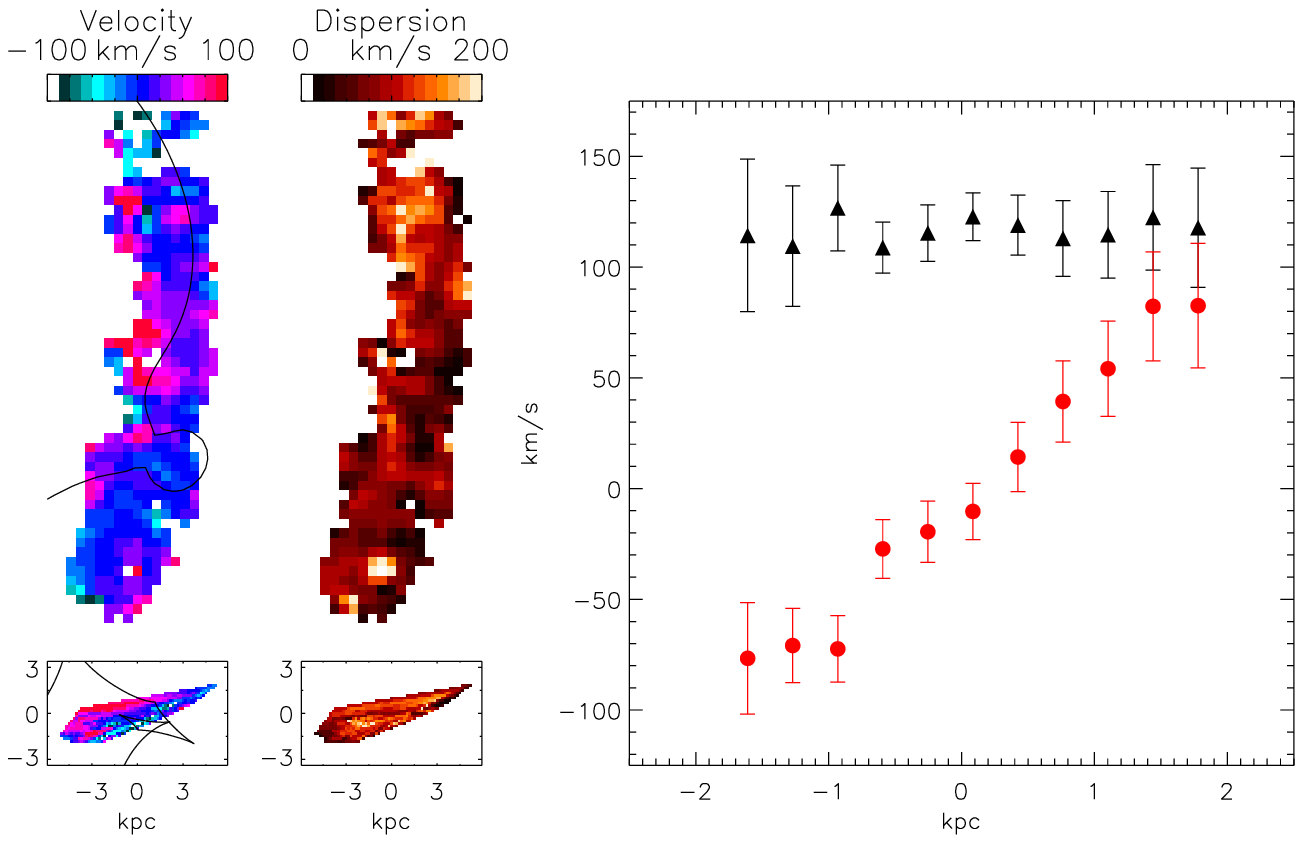}
\caption{\label{fig:kin} Kinematics of the Clone. Left: velocity and velocity dispersion maps in the image plane (top) and 
source plane (bottom); North is up and East is to the left. The critical line and corresponding caustic are overlaid on the velocity maps. Right: one-dimensional velocity (circles) and dispersion (triangles) profiles 
in the source plane extracted along a slit passing through the centroid of H$\alpha$ emission and oriented to maximize 
the velocity gradient. The same velocity gradient is seen in both images (A3-4) of the galaxy, with redshifted 
gas at the center of the IFU and blueshifted gas at both edges of the field.}
\end{figure*}

Considering the kinematic data for H$\alpha$ in the image plane where the instrument resolution and sensitivity are nearly 
uniform and easily characterized, the internal properties of the Clone are fairly typical of those $z\simeq$2-3
galaxies for which data of comparable resolution is available (e.g. \citealt{Stark08, Jones10}).  After smoothing with a 
Gaussian kernel of FWHM $= 0.25$ arcsec we determine the H$\alpha$ emission line velocity and width from a Gaussian fit to the line profile. The intrinsic velocity 
dispersion is determined by subtracting the instrumental resolution of $\sigma_{inst} = 52$ km s$^{-1}$ (measured from  OH sky lines) 
in quadrature from the best-fit line width. Only fits with signal-to-noise $> 5$ in H$\alpha$ are kept. 

In Figure~\ref{fig:kin} we present source-plane velocity and dispersion maps constructed from the A3 image
(see Figure~\ref{fig:cl}). The 2D map and 1D velocity profile both show a large gradient of $\Delta v = \pm 80  \pm 18$ km s$^{-1}$
across the galaxy with some degree of asymmetry. The reconstruction of images A4-5 gives a consistent result with $\Delta 
v = \pm 86 \pm 19$ km s$^{-1}$. The velocity dispersion is typically $\sigma = 130 \pm 20$ km/s in the center and shows similar
asymmetry. The mean dispersion of individual pixels is $\bar{\sigma} = 85 \pm 2$ km s$^{-1}$, consistent with  $\sigma = 80 \pm 4$ km s$^{-1}$ 
derived from the \cite{Hainline09} integrated spectrum. This dispersion gives a dynamical mass $M_{dyn} = 5 r_h \bar{\sigma}^2 / G = 2.4 \times 10^{10}$ M$_{\odot}$ for $r_h = 2.9$ kpc \citep{Hainline09}.

The above observations are characteristic features of star-forming galaxies observed at $z \sim 2$ (e.g.\ the SINS sample
discussed by \citealt{Forster09}). The SINS galaxies have a median dispersion of $\sigma_{int} = 130$ km s$^{-1}$
and $\Delta v / \sigma_{int} = 0.4-2.0$ for those classified as likely disk progenitors. Even those SINS galaxies with the most 
prominent rotational motion show asymmetric features, with model residuals of $\gtrsim 100$ km/s in both velocity and dispersion 
\citep{Cresci09}. 

\section{Gas Phase Metallicity Measurements}

\begin{figure}
\includegraphics[width=\columnwidth]{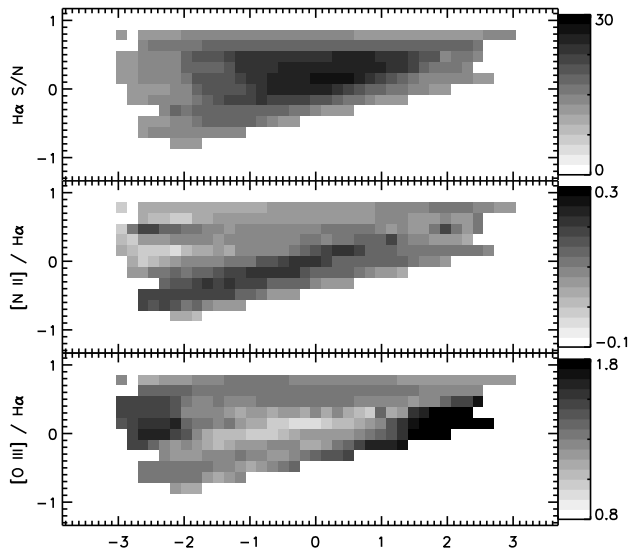}
\caption{\label{fig:source} Source plane reconstruction of the H$\alpha$ signal-to-noise, [N{\sc ii}]/H$\alpha$, and [O{\sc iii}]/H$\alpha$ maps from the A3 region of the arc. Axes are in kpc. The maps show a maximum [N{\sc ii}]/H$\alpha$ and minimum [O{\sc iii}]/H$\alpha$ near the center of the galaxy, slightly offset from the H$\alpha$ emission peak.}
\end{figure}

Reliable metallicity measurements require knowledge of the electron temperature, ionization state, and reddening, for which several emission 
lines must be observed. In practice, all metallicity measurements at high redshift rely on strong emission lines which can be detected in a 
reasonable integration time. Ratios of strong emission lines are then used to infer the gas-phase oxygen abundance, calibrated locally 
using either direct measurements of the electron temperature or via photoionization models. One of the more popular diagnostics is the 
[N{\sc ii}]$\lambda6584$/H$\alpha$ ratio (N2) which \cite{Pettini04} showed is strongly correlated with the oxygen abundance via:
$$12 + \log{O/H} = 8.90 + 0.57 \times \log{ \mbox{[N{\sc ii}]/H}\alpha }$$
with an intrinsic 1$\sigma$ scatter of 0.18 dex.

The main drawbacks of the N2 method are that [N{\sc ii}]/H$\alpha$ does not necessarily trace the global metallicity when (i) AGN and 
shock excitation contribute to [N{\sc ii}] emission; (ii) secondary production of nitrogen leads to variation in the N/O ratio; and (iii) [N{\sc ii}] 
cooling saturates (at $12+\log{O/H} \gtrsim 9.0$). The advantage of N2 lies in the proximity of the two required emission lines, 
such that systematic errors due to reddening and instrumental effects are negligible.

Fortunately, we can use [O{\sc iii}]$\lambda5007$ as an additional constraint on shock ionization, AGN, and variations in the N/O ratio.
We smoothed the data cube spatially with a $0.4$ arcsec FWHM Gaussian, providing a signal-to-noise $\gtrsim 3$ for [N{\sc ii}] across the
entire OSIRIS field of view. We then fit a Gaussian profile to the H$\alpha$ line and determined the best-fit [NII] and [OIII] fluxes for the same profile, 
requiring the H$\alpha$ signal-to-noise to be $> 10$ (Figure~\ref{fig:cl}). We found no significant offset in the velocity 
when fitting these lines independently. For each pixel we infer the [O{\sc iii}]$\lambda5007$/H$\beta$ ratio (assuming Case B recombination) 
and measure the [N{\sc ii}]/H$\alpha$ ratio adopting
an average selective extinction of E(B-V) $= 0.28$ \citep{Hainline09}. The resulting data show no evidence of shock ionization or AGN in the 
standard BPT diagnostic diagram \citep{Baldwin81}. We can examine possible variations in the N/O abundance using the photoionization 
models presented in \cite{Kewley02}. The [O{\sc ii}]/[O{\sc iii}] flux from \cite{Hainline09}, corrected for reddening via the 
H$\alpha$/H$\gamma$ ratio, suggests an ionization parameter $\log(U) = -2.9$ for the integrated spectrum. We compute the metallicity of each 
pixel from both the [N{\sc ii}]/H$\alpha$ and [N{\sc ii}]/[O{\sc iii}] ratios using this ionization parameter and find excellent agreement, with an average 
offset 0.02 and 1$\sigma$ scatter of $0.08$ in $12 + \log{O/H}$. 

\begin{figure}
\includegraphics[width=\columnwidth]{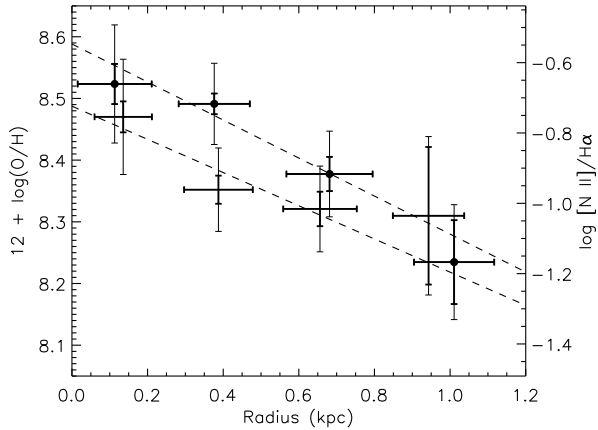}
\caption{\label{fig:n2} Intrinsic metallicity gradient of the Clone arc derived using the N2 index. Measured values of [N{\sc ii}]/H$\alpha$ are shown on the right axis with inferred oxygen abundance on the left. Filled circles and crosses correspond to the source plane properties of images A4 and A3, respectively. The dashed lines are linear fits to the two data sets. Thick error bars represent only the statistical uncertainty in [N{\sc ii}]/H$\alpha$ for each bin, while thin error bars include an additional uncertainty from the N2 index calibration as discussed in the text.}
\end{figure}

The source-plane reconstruction is done independently for the A3 and A4/A5 regions, yielding two independent measurements of the 
metallicity distribution. The results are shown in Figures~\ref{fig:source} and \ref{fig:n2}. [N{\sc ii}]/H$\alpha$ peaks in the center of the galaxy roughly coincident with the global minimum 
of [O{\sc iii}]/H$\alpha$, with [N{\sc ii}]/H$\alpha$ decreasing and [O{\sc iii}]/H$\alpha$ increasing at larger radii. This is the expected signature 
of a metallicity gradient as seen in local disk galaxies (e.g. \citealt{Magrini07}). Using \cite{Pettini04}'s calibration to convert the measured [N{\sc ii}]/H$\alpha$ to an inferred metallicity $12 + \log{O/H}$, we
extract pixels within a slit oriented along the direction of highest magnification to maximize spatial sampling and resolution. We bin pixels 
within the extraction slit based on their radius from the center of the galaxy, determined from the centroid of the global minimum [O{\sc iii}]/H$\alpha$.
 A linear fit to each set of data gives $\frac{d\log{O/H}}{dR} = -0.31 \pm 0.07$ and $-0.27 \pm 0.08$ dex/kpc, with central values of 
$12 + \log{O/H} = 8.59 \pm 0.04$ and $8.49 \pm 0.04$ respectively. Taken together, we infer a metallicity gradient of $\frac{d\log{O/H}}{dR} 
= -0.27 \pm 0.05$ dex/kpc. Varying the central position by up to 800 parsec results in gradients typically shallower by $0.05 \pm 0.05$ dex/kpc, but a
consistent central metallicity. The metallicity inferred for the global spectrum is $12 + \log{O/H} = 8.42 \pm 0.18$ (adopting the scatter in the N2 calibration), consistent
with \cite{Hainline09} within the uncertainties. 

The preceding analysis assumes that scatter in the N2 calibration applies to individual measurements but not their differences. 
We also measured the metallicity gradient with scatter in the N2 calibration treated as an additional uncertainty in each measurement (and their differences), but with no overall systematic offset. The resulting central metallicity is $12 + \log{O/H} = 8.53 \pm 0.06$ and the gradient is $\frac{d\log{O/H}}{dR} = -0.27 \pm 0.11$ dex/kpc. As discussed above, the [O{\sc iii}] measurements suggest no significant error in the inferred metallicity from varying ionization parameter and N/O ratio, implying that scatter in the N2 calibration does not affect relative abundances such as gradients. This is supported by studies of nearby galaxies \citep{Rupke10, Kewley10}, hence we adopt an uncertainty of $\pm 0.05$ dex/kpc in the metallicity gradient. Regardless, Figure~\ref{fig:n2} clearly shows a significant radial gradient with correlation coefficients of $-0.98$ and $-0.89$ for images A4 and A3 of the arc, respectively.

\section{Discussion}

The Clone arc is evidently representative of UV-bright star-forming galaxies at $z\simeq2$ both in its integrated star formation rate of $\simeq\,50 M_\odot$ yr$^{-1}$
and its resolved kinematics. Via the detailed study of 3 emission lines, we have found a strong radial gradient in both the [N{\sc ii}]/H$\alpha$ and [O{\sc iii}]/H$\alpha$ 
flux ratios, consistent with a metallicity gradient which we measure to be $\frac{d\log{O/H}}{dR} = -0.27 \pm 0.05$ dex/kpc. This is stronger than that  
in local disk galaxies, which typically have gradients of $\simeq -0.05$ dex/kpc \citep{Bresolin10, Moran10, Rupke10, Carrera08, Magrini07}, although the total range $\Delta\log{O/H} \simeq 0.5$ is similar. The near-solar central metallicity $12 + \log{O/H} = 8.5 \pm 0.2$ is comparable though somewhat lower than in local massive disk galaxies. We can place a rough upper limit on the age of the Clone by considering the timescale $\tau_{SF} = \frac{M_*}{SFR} < \frac{M_{dyn}}{SFR} = 500$ Myr, suggesting that strong central enhancement occurs very rapidly.

Negative metallicity gradients suggest that the inner regions are older and more gas-depleted than the outskirts. The extremely steep gradient and similar range of 
metallicity in such a young galaxy as the Clone suggests an inside-out growth scenario in which the center is enriched rapidly while stellar mass and metallicity gradually increase at
large radii. This is further supported by kinematic evidence: the large velocity dispersion should cause the metallicity gradient in the Clone to flatten due to radial 
mixing on timescales of order $\tau_{dyn} = d/\bar{\sigma} = 70$ Myr. Infall of metal-poor gas onto the outskirts of the galaxy is therefore necessary to maintain a strong 
gas-phase metallicity gradient, which will naturally flatten toward values observed in local disks when accretion stops.

The chemical and dynamical properties of the Clone are consistent with cosmological simulations which show that accretion of cold gas onto a turbulent disk leads 
to a bulge plus thin disk morphology \citep{Ceverino10, Brooks09}. The ``clumpy disk'' phase can last for several $10^8$ years, during which radial 
abundance gradients would be affected by the competing processes of radial mixing (which flattens the gradient) 
and infall of metal-poor gas (which steepens the gradient), ultimately keeping the abundance spread relatively similar. Meanwhile the galaxy grows in size and 
stellar mass, while centrally-migrating clumps form a bulge which stabilizes the disk.

As emphasized in \cite{Jones10} our  results demonstrate the considerable advantage of using lensed galaxies to probe the internal structure of $z\simeq$2-3
galaxies, both in terms of increased flux {\it and} source plane resolution. Procuring a larger sample of highly magnified $z>$2 galaxies is thus a highly profitable exercise
for furthering our understanding of galaxy assembly and metal enrichment in the Universe.

\section*{Acknowledgments}
EJ is supported by the NPP, administered by Oak Ridge Associated
Universities through a contract with NASA. Part of this work was
carried out at Jet Propulsion Laboratories, California Institute of
Technology under a contract with NASA. JR acknowledges support from
an EU-Marie Curie fellowship.
We thank the referee for useful comments which improved this paper.
TAJ thanks Wal Sargent and Judy Cohen for helpful discussions. We thank the Keck Observatory staff for their assistance in obtaining these observations. The authors wish to recognize and acknowledge the very significant cultural role and reverence that the summit of Mauna Kea has always had within the indigenous Hawaiian community.  We are most fortunate to have the opportunity to conduct observations from this mountain.


\end{document}